# On-chip coherent frequency-domain THz spectroscopy for electrical transport

Katsumasa Yoshioka[*], Norio Kumada, Koji Muraki, and Masayuki Hashisaka

NTT Basic Research Laboratories, NTT Corporation, 3-1 Morinosato-Wakamiya, Atsugi, 243-0198, Japan, E-mail: katsumasa.yoshioka.ch@hco.ntt.co.jp

**Abstract**: We developed a coherent frequency-domain THz spectroscopic technique on a coplanar waveguide in the ultrabroad frequency range from 200 MHz to 1.6 THz based on continuous wave (CW) laser spectroscopy. Optical beating created by mixing two frequency-tunable CW lasers is focused on photoconductive switches to generate and detect high-frequency current in a THz circuit. In contrast to time-domain spectroscopy, our frequency-domain spectroscopy enables unprecedented frequency resolution of 10 MHz without using complex building blocks of femtosecond laser optics. Furthermore, due to the coherent nature of the photomixing technique, we are able to identify the origin of multiple reflections in the time domain using the Hilbert analysis and inverse Fourier transform. These results demonstrate that the advantages of on-chip coherent frequency-domain spectroscopy, such as its broadband, frequency resolution, usability, and time-domain accessibility, provide a unique capability for measuring ultrafast electron transport in integrated THz circuits.

High-frequency electrical transport measurement in the gigahertz (GHz) frequency range using radio-frequency (RF) circuits is a crucial technology for investigating and controlling electron dynamics in various systems [1]. Both time-domain spectroscopy (TDS) [2–8] and frequency-domain spectroscopy (FDS) [4,9–13] have been used as complementary techniques. For example, they have been used to unveil nontrivial dynamics of collective and single-particle excitations in two-dimensional (2D) electron systems and one-dimensional edge channels in the quantum Hall regime. However, applying these RF measurements to study emergent phenomena in quantum nanocircuits [14,15] and materials such as topological insulators [16] and exfoliated 2D crystals [17] remains a challenge. This is because the coherent length or sample size in these systems is on the order of 10 μm; for the typical charge velocity of ~$10^6$ m/s and bandwidth of ~10 GHz (corresponding time resolution of ~100 ps), a distance of at least ~100 μm is necessary to resolve charge transport. Hence, increasing the measurement bandwidth to the terahertz (THz) range is highly desirable for probing novel transport phenomena in such circuits and materials, such as, ultrafast quantum interference [14,15], charge transport in helical edge channels [18,19], and dynamical signatures of Majorana zero modes [20].

On-chip THz spectroscopy can overcome the issues of frequency and sample size. On-chip generation and detection of sub-picosecond electrical pulses in the time domain have been achieved using a THz circuit and femtosecond laser technologies [21–27]. Recently, on-chip THz-TDS has been applied to investigate ultrafast electron transport in graphene [28–30] and semiconductor 2D electron gas [31] by integrating them on the THz circuit. Though an extremely powerful technique, THz-TDS has some limitations: its frequency resolution is typically tens of GHz limited by the tunable optical path length of the delay stage, and thus accessing the GHz range with THz-TDS is challenging. In addition, its complicated and costly optical setup based on ultrafast femtosecond laser systems requires high expertise, thus limiting its use to specialized groups. On the other hand, FDS has advantages in terms of bandwidth, frequency resolution, and usability. Phase-sensitive (or coherent) FDS with time-domain accessibility will accelerate the ongoing progress of ultrafast transport studies if implemented on-chip. However, unlike in RF measurements, coherent detection of electrical current by frequency-domain measurements is difficult in optics, which has left the possibility of on-chip THz-FDS largely unexplored.

Here, we demonstrate on-chip coherent THz-FDS by applying photomixing techniques [32,33] and optimizing the THz circuit. We achieve an ultrabroad frequency range from 200 MHz to 1.6 THz and an unprecedented frequency resolution of 10 MHz. The coherent nature of the photomixing technique allows us to extract time-domain dynamics from the frequency-domain signal using the Hilbert analysis and inverse Fourier transform [33]. These advantages, together with the cost-effective and straightforward optical setup of our on-chip THz-FDS, will make ultrafast electron transport phenomena more accessible in various circuits and materials.

Figure 1(a) shows a schematic of our THz circuit. The circuit consists of a coplanar waveguide (CPW) and two PC switches (PC1 and PC2) attached to it. Two types of THz circuits with/without a DC block were examined. These circuits were fabricated on a low-temperature-grown GaAs substrate (BATOP GmbH) by optical lithography and evaporation of Ti/Au (10 nm/180 nm). The 400-µm-long CPW transmission region consists of a 30-µm-wide center conductor and two ground planes separated from the center conductor by 20 µm. As shown in Fig.1(b), frequency-tunable optical beating is generated by mixing two distributed-feedback diode lasers (TOPTICA Photonics, TeraScan 780), which is then divided into pump and probe beams with the same average power of ~20 mW. The two orthogonally polarized beams are combined using a polarization beam splitter, aligned with a slight displacement to focus them onto PC1 and PC2 with an objective lens. PC1 is biased with a DC voltage and excited by the pump beam to generate a THz wave, which propagates through the CPW and is detected as a THz current flowing through PC2 excited by the probe beam. An optical chopper modulates the pump beam at a few hundred hertz for lock-in detection of the THz current. The detected THz current $I_{THz}$ is determined by the interference between the incoming THz signal and optical beating at PC2. As a result, $I_{THz}$ is given by

$$I_{THz} \propto E_{THz}\cos(\Delta\varphi) = E_{THz}\cos(2\pi\Delta L f_{THz}/c), \qquad (1)$$

where $E_{THz}$ is the THz amplitude, $\Delta\varphi$ is the phase difference between the THz signal and the optical beating, $f_{THz}$ is the THz frequency, $c$ is the speed of light, and $\Delta L$ is the optical path difference. $\Delta L$ is represented by $\Delta L = L_{pump} - L_{probe} + L_{sample}$, where $L_{pump}$ and $L_{probe}$ are the optical path lengths of the pump and probe beams, respectively, and $L_{sample}$ is the propagation distance of the THz current. By scanning $f_{THz}$, we can obtain the phase $\Delta\varphi(f_{THz})$ as well as amplitude $E_{THz}(f_{THz})$ spectra of the THz current.

Figure 2 shows the measured $I_{\text{THz}}$ as a function of $f_{\text{THz}}$ for the THz circuit without a DC block. We chose the bias voltage for PC1 in such a way that the signal-to-noise ratio was maximized for each THz circuit. $I_{\text{THz}}$ shows oscillation as expected from Eq. (1). The decreasing trend in the oscillation amplitude with increasing $f_{\text{THz}}$ is due to the finite relaxation time (~400 fs [34]) of photoexcited carriers in the PC switches. The finite offset (~2 nA) comes from the leak current of the PC switches. The close-up of the low-frequency regime (0–2 GHz) shows that clear oscillations persist down to 200 MHz, demonstrating the capability of THz-FDS to access the sub-GHz regime. The frequency resolution of the amplitude spectrum obtained by the Hilbert analysis [33] (black solid curve) reaches 10 MHz, which in the present case is limited by the frequency scan step throughout the whole spectrum. This unprecedented frequency resolution comes from the narrow linewidth inherent to CW lasers. Note that the offset part of the signal fluctuates and therefore degrades the signal-to-noise ratio, limiting the bandwidth of the THz circuit. We consider that the fluctuation is associated with the stability of the PC switches, which is affected by the CW laser heating and carrier accumulations [35].

To improve the signal-to-noise ratio and achieve higher bandwidth, we put a DC block [36] in the center conductor [Fig. 1(a)]. The DC block is a 3-μm gap, which cuts the signal below 1 GHz. As shown in the inset of Fig. 3, the DC block reduces the unwanted offset to ~0.4 nA. Figure 3 compares the amplitude spectra of THz circuits with/without the DC block. The signal-to-noise ratio is improved by a factor of ~5, and the measurement bandwidth is improved from 0.6 THz (without the DC block) up to 1.6 THz (with the DC block). These results indicate that our method enables ultrabroadband spectroscopy ranging from RF (sub-GHz) to optical technology (THz) domains with unprecedented frequency resolution, which is not available in either RF technology or on-chip THz-TDS. Note that the dips in the amplitude spectrum at around 0.8 and 1.0 THz are due to the interference between even and odd modes [37] and are outside the focus of this study.

To use FDS to study electron transport in various samples and elucidate the ultrafast dynamics therein, it is crucial to access the time-domain signal. Our phase-sensitive measurements allow the time-domain signal to be extracted without any of the uncertainty that inevitably arises from K-K transformation in conventional incoherent THz-FDS. Figure 4(a) shows the time-domain signal obtained by the Hilbert analysis and inverse Fourier transform [33]. As shown in the inset, the length of the whole time-domain signal is 100 ns, which corresponds to the inverse of the frequency scan step of 10 MHz. The signal associated with the THz circuit appears between 5.260 and 6.512 ns, which we plot in a magnified view in Fig. 4 as a function of the time difference $\Delta t$ from the first peak. The inverse Fourier transform of the time-domain signal for the window $-0.089 \leq \Delta t \leq 1.163$ ns [red trace in Fig. 4(b)] shows less noise than the original amplitude spectrum (gray trace). This indicates that we can remove unwanted interferences from optical components, such as the lenses, connectors, and beam splitter, by selecting the signal in the time domain.

Further careful analysis of the time-domain signal enables us to identify the origin of the Fabry-Pérot (FP) effect, which appears as minor oscillations superimposed on the amplitude spectrum. When the window for the inverse Fourier transform is further narrowed to $-0.089 \leq \Delta t \leq 0.054$ ns, the FP effect disappears [black trace in Fig. 4(b)]. Therefore, the negative peak after $\Delta t = 0.054$ ns (highlighted by the blue circle) can be identified as the first reflection that causes the FP effect. By considering the propagation time $\Delta t = 60$ ps and the propagation velocity $1.2 \times 10^8$ m/s [38] of the THz current, the associated propagation length is determined to be 7.2 mm. This long propagation length suggests back reflection from the contact pad located at the end of the coplanar waveguide 7.1 mm away. We note that the detailed analysis of the time-domain signal presented here is a powerful technique to investigate the dynamics of ultrafast electron transport of any material — it is not limited to the FP effect in the THz circuit.

In summary, we reported an on-chip coherent frequency-domain THz spectroscopic technique in the ultrabroad frequency range from 200 MHz to 1.6 THz. Its high frequency resolution of 10 MHz and capability to access the time-domain signal is highly applicable to investigate ultrafast transport phenomena in quantum circuits and materials by integrating them on the THz circuit. The accessibility of the sub-GHz and THz regime enables a comprehensive study of slow and ultrafast dynamics with high reproducibility using a single experimental setup. We believe that the ongoing progress of ultrafast transport studies, which currently require high expertise in ultrafast femtosecond optics, will be significantly accelerated by the cost-effective and straightforward optical setup of our technique.

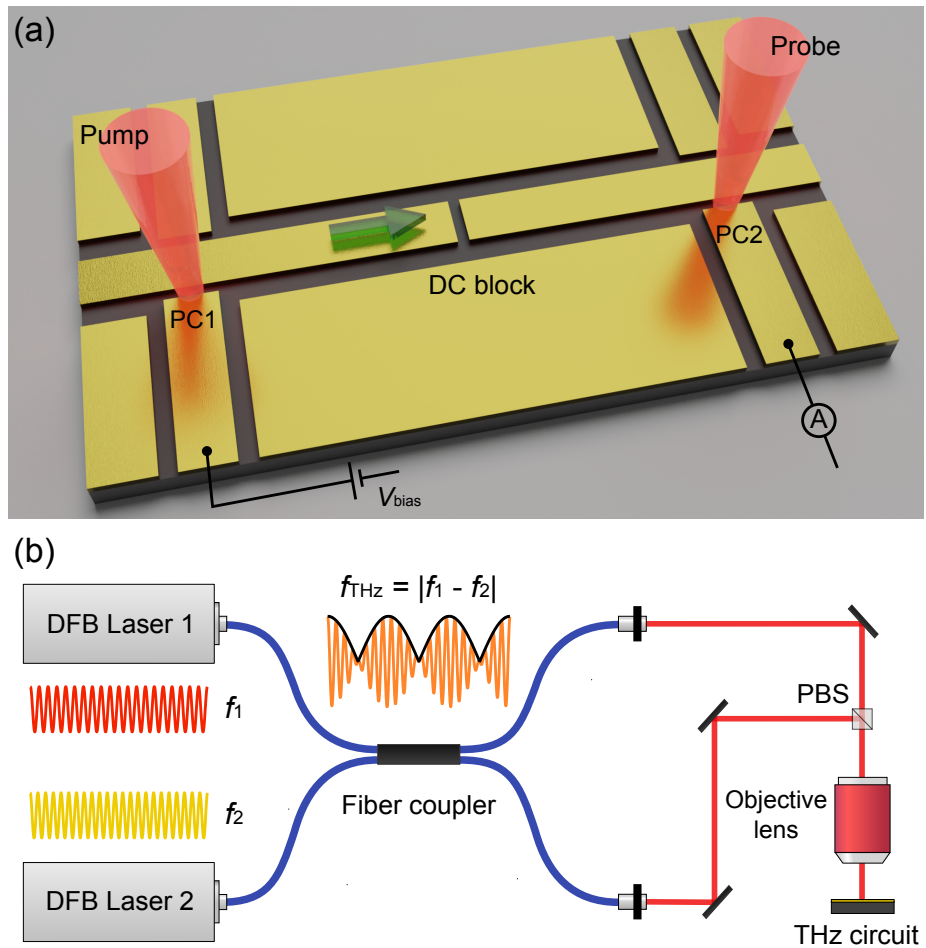

FIG. 1. (a) Illustration of the on-chip THz circuit. The circuit consists of photoconductive switches (PC1 and PC2), a coplanar waveguide, and a DC block. Two types of THz circuit with/without the DC block were investigated. Beating laser beams are focused on the PC switches to generate and detect the THz current in a coherent manner. All conductors except PC switches are connected to the ground. (b) Schematic of the experimental setup. DFB Laser, distributed-feedback diode laser; PBS, polarization beam splitter. The frequency of THz current $f_{THz}$ is controlled by the frequency of the optical beating created by mixing DFB laser 1 and 2.

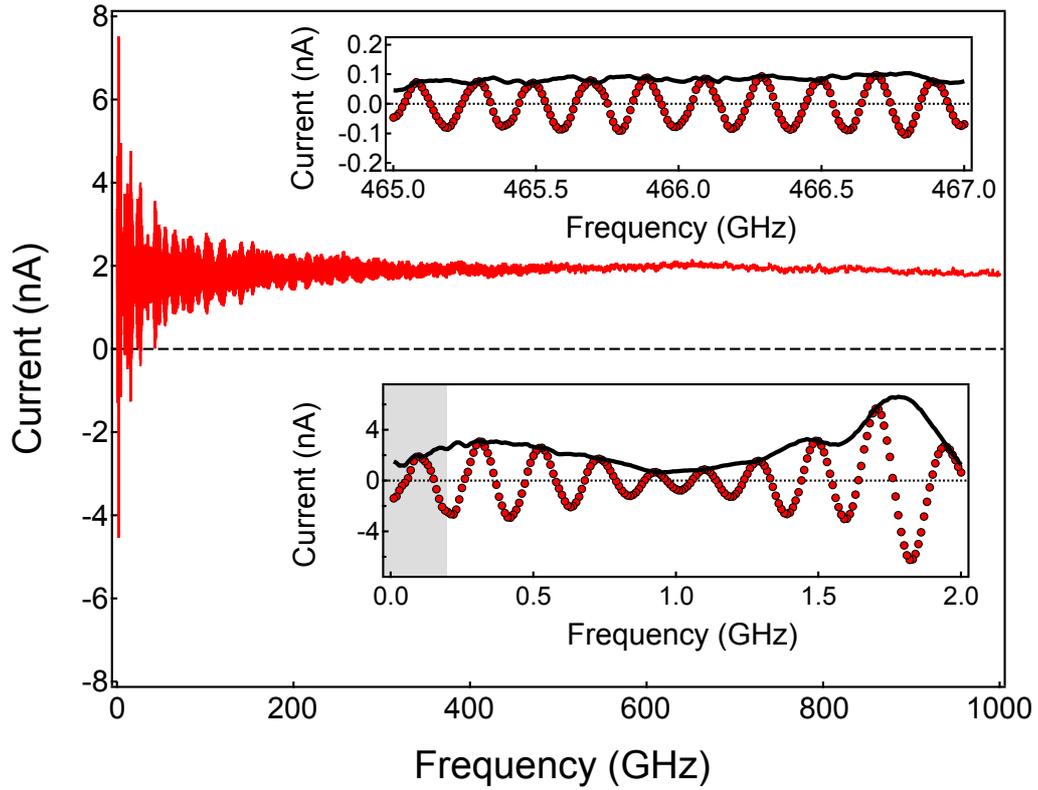

**FIG. 2.** THz current as a function of frequency measured in the THz circuit without the DC block. Applied DC voltage is $V_{bias}$ = 20 V. The insets show close-ups of the low- and high-frequency ranges after subtracting the offset. The gray shaded area (0–200 MHz) represents the regime where the oscillatory signal exhibits slight distortion. The envelope, which represents the amplitude spectrum, was obtained using the Hilbert analysis.

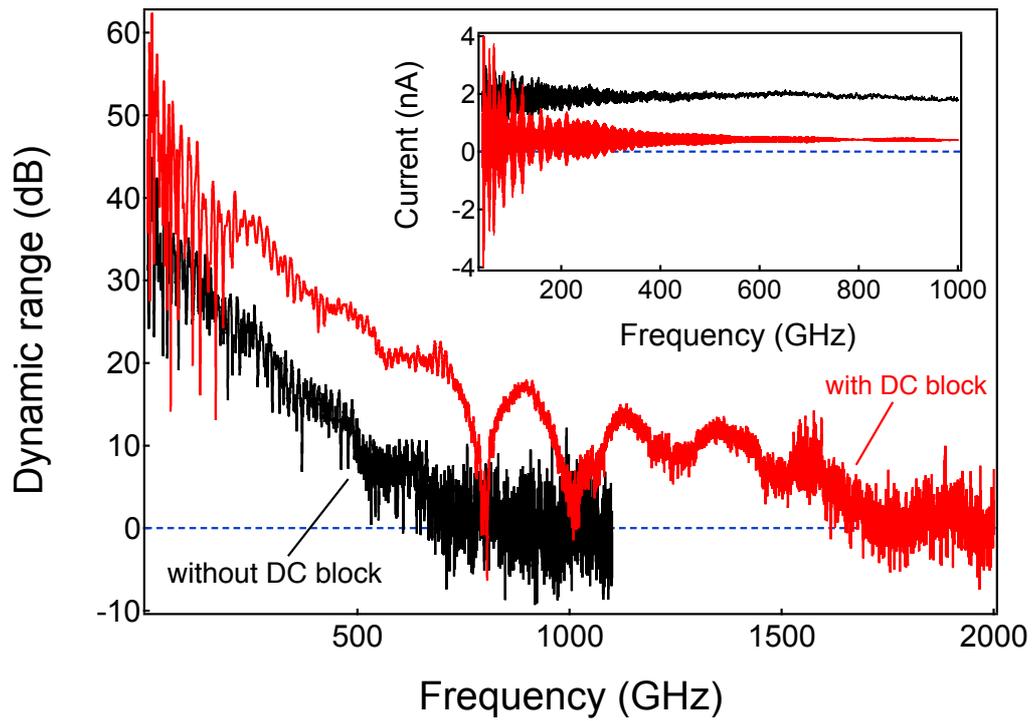

**FIG. 3.** Amplitude spectra of the THz circuits with/without the DC block. Applied DC voltage $V_{bias}$ is 20 and 30 V for the circuits without and with the DC block, respectively. The inset shows measured THz current.

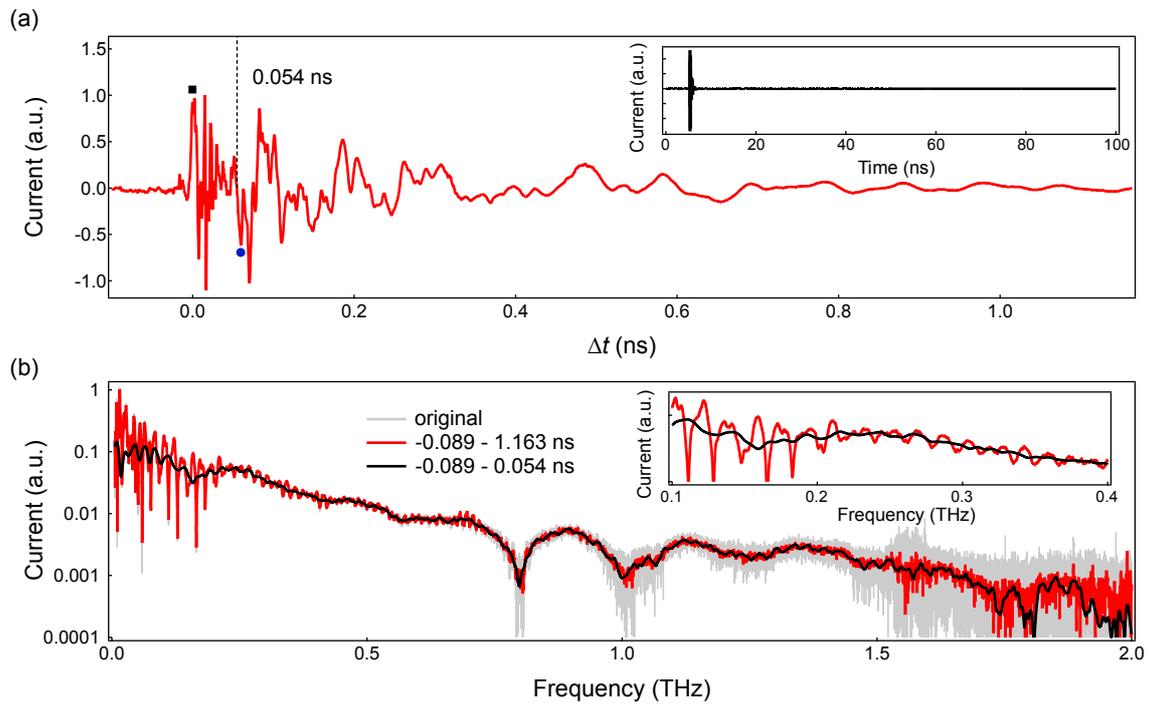

**FIG. 4.** (a) Time-domain signal obtained using the Hilbert analysis and inverse Fourier transform. The first peak marked by the black square shows the beginning of the THz signal, while the blue circle marks the beginning of signals due to the reflections that cause the Fabry-Pérot effect. The inset shows the whole time-domain signal. (b) Fourier-transformed spectra of (a) for different time windows. The inset shows a close-up from 0.1 to 0.4 THz.

**Acknowledgement**

This work was supported by a Grants-in-Aid for Scientific Research (Grant No. JP17K18750, JP16H06009). We thank H. Murofushi for technical support.

The data that support the findings of this study are available from the corresponding author upon reasonable request.

**References**

[1] C. Bäuerle, D. Christian Glattli, T. Meunier, F. Portier, P. Roche, P. Roulleau, S. Takada, and X. Waintal, Reports Prog. Phys. **81**, 056503 (2018).

[2] R. C. Ashoori, H. L. Stormer, L. N. Pfeiffer, K. W. Baldwin, and K. West, Phys. Rev. B **45**, 3894 (1992).

[3] G. Ernst, N. B. Zhitenev, R. J. Haug, and K. von Klitzing, Phys. Rev. Lett. **79**, 3748 (1997).

[4] N. Kumada, P. Roulleau, B. Roche, M. Hashisaka, H. Hibino, I. Petković, and D. C. Glattli, Phys. Rev. Lett. **113**, 266601 (2014).

[5] H. Kamata, N. Kumada, M. Hashisaka, K. Muraki, and T. Fujisawa, Nat. Nanotechnol. **9**, 177 (2014).

[6] M. Hashisaka, N. Hiyama, T. Akiho, K. Muraki, and T. Fujisawa, Nat. Phys. **13**, 559 (2017).

[7] G. Roussely, E. Arrighi, G. Georgiou, S. Takada, M. Schalk, M. Urdampilleta, A. Ludwig, A. D. Wieck, P. Armagnat, T. Kloss, X. Waintal, T. Meunier, and C. Bäuerle, Nat. Commun. **9**, 2811 (2018).

[8] M. Kataoka, N. Johnson, C. Emary, P. See, J. P. Griffiths, G. A. C. Jones, I. Farrer, D. A. Ritchie, M. Pepper, and T. J. B. M. Janssen, Phys. Rev. Lett. **116**, 126803 (2016).

[9] M. Hashisaka, H. Kamata, N. Kumada, K. Washio, R. Murata, K. Muraki, and T. Fujisawa, Phys. Rev. B **88**, 235409 (2013).

[10] E. Bocquillon, V. Freulon, J.-M. Berroir, P. Degiovanni, B. Plaçais, A. Cavanna, Y. Jin, and G. Fève, Nat. Commun. **4**, 1839 (2013).

[11] A. C. Mahoney, J. I. Colless, L. Peeters, S. J. Pauka, E. J. Fox, X. Kou, L. Pan, K. L. Wang, D. Goldhaber-Gordon, and D. J. Reilly, Nat. Commun. **8**, 1836 (2017).

[12] A. C. Mahoney, J. I. Colless, S. J. Pauka, J. M. Hornibrook, J. D. Watson, G. C. Gardner, M. J. Manfra, A. C. Doherty, and D. J. Reilly, Phys. Rev. X **7**, 011007 (2017).

[13] C. Lin, K. Morita, K. Muraki, and T. Fujisawa, Jpn. J. Appl. Phys. **57**, (2018).

[14] B. Gaury and X. Waintal, Nat. Commun. **5**, 3844 (2014).

[15] B. Gaury, J. Weston, and X. Waintal, Nat. Commun. **6**, 6524 (2015).

[16] M. König, H. Buhmann, L. W. Molenkamp, T. Hughes, C. X. Liu, X. L. Qi, and

S. C. Zhang, J. Phys. Soc. Japan **77**, 1 (2008).

[17] A. K. Geim and I. V Grigorieva, Nature **499**, 419 (2013).

[18] A. Calzona, M. Carrega, G. Dolcetto, and M. Sassetti, Phys. Rev. B **92**, 195414 (2015).

[19] F. Dolcini, R. C. Iotti, A. Montorsi, and F. Rossi, Phys. Rev. B **94**, 165412 (2016).

[20] R. Tuovinen, E. Perfetto, R. van Leeuwen, G. Stefanucci, and M. A. Sentef, New J. Phys. **21**, 103038 (2019).

[21] M. B. Ketchen, D. Grischkowsky, T. C. Chen, C.-C. Chi, I. N. Duling, N. J. Halas, J.-M. Halbout, J. A. Kash, and G. P. Li, Appl. Phys. Lett. **48**, 751 (1986).

[22] C. Wood, J. Cunningham, P. C. Upadhya, E. H. Linfield, I. C. Hunter, A. G. Davies, and M. Missous, Appl. Phys. Lett. **88**, 2004 (2006).

[23] C. Russell, C. D. Wood, A. D. Burnett, L. Li, E. H. Linfield, A. G. Davies, and J. E. Cunningham, Lab Chip **13**, 4065 (2013).

[24] N. Hunter, A. S. Mayorov, C. D. Wood, C. Russell, L. Li, E. H. Linfield, A. G. Davies, and J. E. Cunningham, Nano Lett. **15**, 1591 (2015).

[25] C. Kastl, C. Karnetzky, H. Karl, and A. W. Holleitner, Nat. Commun. **6**, 6617 (2015).

[26] C. Karnetzky, P. Zimmermann, C. Trummer, C. Duque Sierra, M. Wörle, R. Kienberger, and A. Holleitner, Nat. Commun. **9**, 2471 (2018).

[27] Y. Yang, R. B. Wilson, J. Gorchon, C.-H. Lambert, S. Salahuddin, and J. Bokor, Sci. Adv. **3**, e1603117 (2017).

[28] P. Gallagher, C. S. Yang, T. Lyu, F. Tian, R. Kou, H. Zhang, K. Watanabe, T. Taniguchi, and F. Wang, Science **364**, 158 (2019).

[29] J. W. McIver, B. Schulte, F.-U. Stein, T. Matsuyama, G. Jotzu, G. Meier, and A. Cavalleri, Nat. Phys. **16**, 38 (2020).

[30] J. O. Island, P. Kissin, J. Schalch, X. Cui, S. R. Ul Haque, A. Potts, T. Taniguchi, K. Watanabe, R. D. Averitt, and A. F. Young, Appl. Phys. Lett. **116**, 161104 (2020).

[31] J. Wu, O. Sydoruk, A. S. Mayorov, C. D. Wood, D. Mistry, L. Li, E. H. Linfield, A. Giles Davies, and J. E. Cunningham, Appl. Phys. Lett. **108**, 091109 (2016).

[32] A. Roggenbuck, H. Schmitz, A. Deninger, I. C. Mayorga, J. Hemberger, R. Güsten, and M. Grüninger, New J. Phys. **12**, 043017 (2010).

[33] D.-Y. Kong, X.-J. Wu, B. Wang, Y. Gao, J. Dai, L. Wang, C.-J. Ruan, and J.-G. Miao, Opt. Express **26**, 17964 (2018).

[34] S. Gupta, M. Y. Frankel, J. A. Valdmanis, J. F. Whitaker, G. A. Mourou, F. W. Smith, and A. R. Calawa, Appl. Phys. Lett. **59**, 3276 (1991).

[35] Y. Minami, K. Horiuchi, K. Masuda, J. Takeda, and I. Katayama, Appl. Phys. Lett. **107**, 171104 (2015).

[36] S. Gevorgian, A. Deleniv, T. Martinsson, S. Gal'chenko, P. Linnér, and I. Vendik, Int. J. Microw. Millimeter-Wave Comput. Eng. **6**, 369 (1996).

[37] I. Wolff, *Coplanar Microwave Integrated Circuits* (John Wiley & Sons, Inc., Hoboken, NJ, USA, 2006).

[38] S. Alexandrou, R. Sobolewski, and T. Y. Hsiang, IEEE J. Quantum Electron. **28**, 2325 (1992).